# Optical transmittance of multilayer graphene

SHOU-EN ZHU[1(a)], SHENGJUN YUAN[2] and G.C.A.M. JANSSEN[1]

[1] *Micro & Nano Engineering Lab, Department of Precision and Microsystems Engineering, Delft University of Technology, Mekelweg 2, 2628CD, Delft, The Netherlands*

[2] *Radboud University of Nijmegen, Institute for Molecules and Materials, Heijendaalseweg 135, 6525AJ, Nijmegen, The Netherlands*

**Abstract** – We study the optical transmittance of multilayer graphene films up to 65 layers thick. By combing large-scale tight-binding simulation and optical measurement on CVD multilayer graphene, the optical transmission through graphene films in the visible region is found to be solely determined by the number of graphene layers. We argue that the optical transmittance measurement is more reliable in the determination of the number of layers than the commonly used Raman Spectroscopy. Moreover, optical transmittance measurement can be applied also to other 2D materials with weak van der Waals interlayer interaction.

**Introduction.** – Graphene is a two-dimensional material with carbon atoms in a honeycomb lattice. It has many potential applications thanks to its unique electrical, mechanical, chemical and optical properties [1-3]. Graphene may outperform existing transparent conductive materials, and a graphene based flexible touch screen was demonstrated by Bae *et al.* in 2010 [4]. Multilayer graphene is a graphene thin film with weak van der Waals interaction between the layers, and its electronic and optical properties are sensitive to the number of layers as well as the stacking sequence [1, 2]. A fast and reliable method to determine the layer number is desired in the fabrication and measurement of multilayer graphene.

For multilayer graphene, Min *et al.* derived that the optical transmission through a graphene films is directly dependent on the optical conductance of the graphene stack, and the optical transmittance $T(\omega)$ of graphene films as a function of incident light frequency $\omega$ can be written as [5, 6]:

$$T(\omega) = \left[1 + \frac{2\pi}{c} G(\omega)\right]^{-2} \quad (1)$$

where $G(\omega)$ is the optical conductivity of the graphene film, and $c$ is the speed of light. In the visible region, by neglecting the interlayer interaction, the optical conductivity of multilayer is linearly proportional to the layer number $N$ as $G(\omega) = NG_1(\omega)$, where $G_1(\omega)$ is the optical conductivity of single layer graphene at frequency $\omega$ [5]. $G_1(\omega)$ only becomes equal to universal optical conductance $G_0 = e^2/(4\hbar)$ in the limit of a massless Dirac fermion bandstructure [2, 7], where $e$ is the elementary charge, and $\hbar$ is the reduced Planck's constant [5, 8]. The optical transmittance of multilayer graphene can be simplified to:

$$T(\omega) = \left[1 + \frac{2\pi}{c} NG_1(\omega)\right]^{-2} \quad (2)$$

where $G_1(\omega) = f(\omega)G_0$. $f(\omega)$ is a correction coefficient to compensate the deviation between $G_1(\omega)$ and $G_0$. The eq. (2) can be further revised including the well defined value $G_0$ as:

$$T(\omega) = (1 + f(\omega)\pi\alpha N/2)^{-2} \quad (3)$$

here, $\alpha = e^2/(\hbar c) \approx 1/137$ is the fine structure constant [9]. Previous work from Nair *et al.* has shown that monolayer graphene can absorb ~2.3% of light. This value can be defined solely by $\pi\alpha$ based on the Dirac cone approximation, which is only valid for the coupling between light and relativistic electrons near the Dirac point [9].

**Numerical simulation.** – In order to obtain more reliable theoretical results of optical conductivity by considering the interlayer hoppings as well as different stacking sequence in multilayer graphene, we carried out the large-scale simulation in the framework of full $\pi$ band tight-binding model. The optical conductivity $G(\omega)$ is calculated numerically by using the Kubo formula [10-12] (omitting Drude weight which is not related to the light adsorption at finite $\omega$),

$$G(\omega) = \lim_{\varepsilon \to +0} \frac{1 - e^{-\beta\hbar\omega}}{\hbar\omega A} \int_0^\infty dt\, e^{i(\omega+i\varepsilon)t} 2i\,\text{Im}\langle\phi|J[1-f(H)]J(t)f(H)|\phi\rangle$$

(4)

where $A$ is the sample area, $\beta = 1/T$ is the inverse temperature, $f(H) = 1/\{\exp[\beta(H-\mu)]+1\}$ is the Fermi-Dirac function of the Hamiltonian operator $H$, $\mu$ is the chemical potential, and $J$ is the current operator. The state $|\phi\rangle$ is a normalized random state which covers all the eigenstates in the whole spectrum. The time evolution operator and Fermi-Dirac operator are represented as the Chebyshev polynomial expansions. In order to phenomenologically implement the red shift of the absorbance spectrum due to the excitonic effect, the hopping energy between two nearest atomic sites is reduced to be $t = 2.3$ eV, the value which leads to the match of the simulated π-excitonic peak at $2t$ and the experimetnal observed peak at 4.6 eV. The numerical method implemented here has the advantage that the CPU time and the memory costs in the simulations are both linear dependent on the size of the sample.

[a]E-mail: shouen.zhu@tudelft.nl

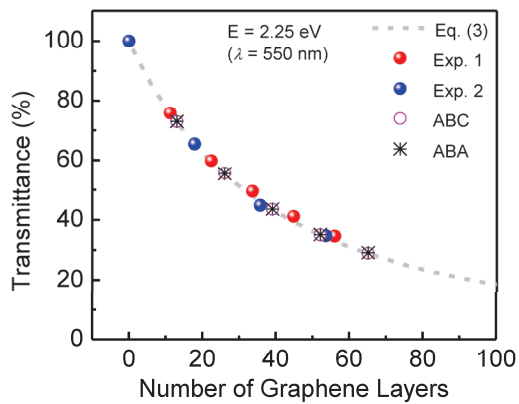

Fig. 1: (Colour on-line) Optical transmittance of CVD multilayer graphene and simulation results. The red circles and blue circles are the experiment data points from multi-stacking. The gray dashed line indicate theory curve from eq. (3). The magenta hollow dots and black stars indicate simulation data from ABC and ABA stacked multilayer graphene films, respectively.

In fig. 1, we plot the numerical results of the optical transmittance of multilayer graphene as a function of the layer number by using the Kubo formula eq. (4). We consider both ABA and ABC stacking sequence. The interlayer hopping parameters between the atomic sites in two nearest layers are set to be $t_1 = 0.12t$, $t_3 = 0.1t$, and $t_4 = -0.04t$ [1, 2]. For incident energy $E = 2.25$ eV ($\lambda = 550$ nm), the absorption of the light is the same for both ABA- and ABC-stacked multilayer graphene, indicating that the optical transmittance of the light at wavelength 550 nm is independent on the stacking sequence. This is due to the fact that interlayer hoppings mainly affect the band strcuture below the energy of $t_1$ and around the van Hove singularities [13]. Furthermore, the numerical results match the analytic approximation expressed in eq. (3), and one can therefore estimate the layer number by measuring the optical transmittance and fit the results to the relation of eq. (3) by using the optical conductivity of single layer graphene at 550nm wavelength.

**Experiment.** – In order to study the optical transmittance experimentally, large area graphene films were synthesized through chemical vapor deposition (CVD). Monolayer graphene and multilayer graphene films were achieved with copper [4] and nickel [14] catalysts, respectively. The monolayer and multilayer graphene films were transferred onto a glass substrate with polymethyl methacrylate (PMMA), followed by etching away the metal catalyst, soaking in deionized (DI) water, releasing the graphene on transparent substrate and solving the PMMA with acetone [4]. The transmittance of monolayer graphene was recorded using a Shimazdu ultraviolet–visible spectrometer (UV-3600). The monolayer CVD graphene shows transmittance of 97.4% at normal incidence for 550 nm wavelength light as shown in fig. 2. This value is slightly lower than 97.7%, which was previously attributed to the polymer residue [4]. However, it is clear that the experimental results of monolayer coincide well with our numerical calculations in the range from 550 nm to 800 nm as shown in fig. 2. Both from the simulations as well as from the tranmittance measurements on the monolayer graphene, we determine the value of $f(\omega)$ to be 1.13 at 550 nm wavelength. As a comparison, the experimental data by Nair *et al.* [9] are plotted in blue hollow circles. These data match slightly less good to our tight-binding simulation. The deviations for $\lambda < 500$ nm can not be reproduced from our numerical simulations by considering small amout of disorder such as carbon vacancies or hydrogen adatoms (data not shown), but might be originated from the excitonic effect which is beyond our phenomenological consideration with a reduced hopping amplitude.

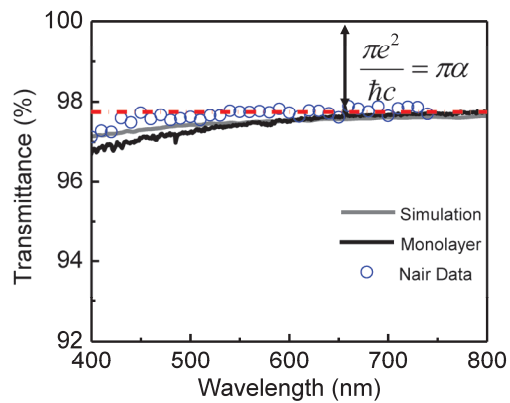

Fig. 2: (Colour on-line) Optical transmittance of CVD monolayer graphene (black line) and simulation results (gray line). The blue open circles are the experiment data points from Nair *et al.* (Reprinted with permission from AAAS). The red dash dot lines indicate the light absorption $\pi\alpha$ of monolayer graphene predicted by Dirac cone approximation.

In order to verify the dependence of the optical transmittance of multilayer graphene layers, two sets of multilayer CVD graphene films were grown on a nickel coated wafer. The numbers of layers in the two sets are determined to be 11.2 and 17.8 by using eq. (3), with $f(\omega) = 1.13$. The multilayer graphene films are polycrystalline with an irregular number of layers, however with uniform optical transparency on a macroscopic scale. These two sets were stacked upon themselves to get multilayers consisting of $n$ times 11.2 and n times 17.8 layers, where $n$ is the number of transfers. The transmittance curves of each of these stacks with $\lambda$ ranging from 400 nm to 800 nm are presented in fig. 3. The numbers in the figures indicate the optical transmittance at 550 nm incident light. We extracted the experimental data at 550 nm (fig. 1) from multi-stacked graphene films in fig. 3. In fig. 1, the red circles represent the stacks fabricated by stacking the 11.2 layer sample, and the blue circles represent the stacks originating from the 17.8 layer sample. The experimental data points coincide very well with eq. (3) as well as our numerical simulations.

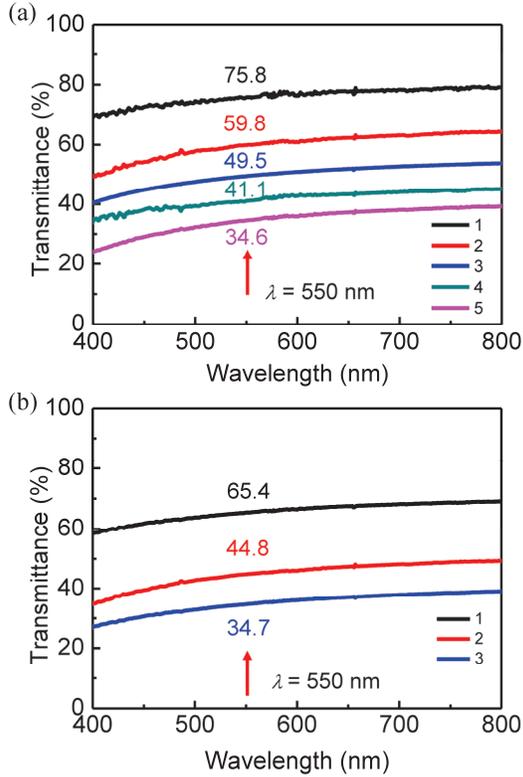

Fig. 3: (Colour on-line) Optical transmittance of graphene films with multi-times transfer in two sets with incident light wavelength ranging from 400 nm to 800 nm. (a) The number of layers is determined to be 11.2. (b) The number of layers is determined to be 17.8. The data in the figures indicate the optical transmittance with incident 550 nm visible light.

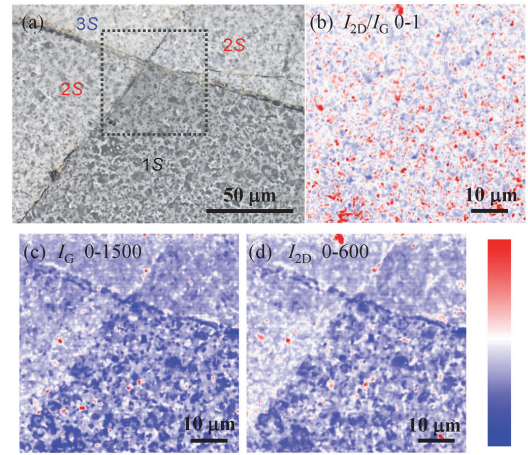

Fig. 4: (Colour on-line) (a) Optical microscope image of multilayer layer graphene films transferred onto glass substrate. 1$S$, 2$S$, and 3$S$ are corresponding to 1, 2, and 3 times stacking regions. (b) 2$D$ and $G$ peak Raman intensity ratio in the overlap multi-stacked region indicated by dotted square in fig. 3a. (c), (d) Raman intensities of $G$ and 2$D$ peak in the same region as in fig.3b.

Raman Spectroscopy has been used to determine the number of layers in multilayer graphene consisting of a few layers [15]. The boundary region with 3 times stacking for 11.2 layers graphene films is presented in fig. 4. The clear step edges for different times stacking can be distinguished in the corresponding region of optical image as shown in fig. 4a. We find that the intensities of Raman $G$ peak and 2$D$ peak increase with stacking graphene films. However, the increase of intensity is no longer distinguishable between twice stacking and three times stacking. Intensity ratio between $G$ mode and 2$D$ mode has been a fingerprint to indentify the number of graphene layers [15]. In contrast, it does not provide clear information for our multi-stacking samples as shown in fig. 4b. Raman spectra data for one time transfer and five times stacking are presented in fig. 5. As number of layers increases, the thickness becomes more uniform. However, no conclusive differences are observed due to the thicker graphene layers in our experiment, which clearly shows the limitation of Raman spectroscopy to determine the number of graphene layers over 9 [16].

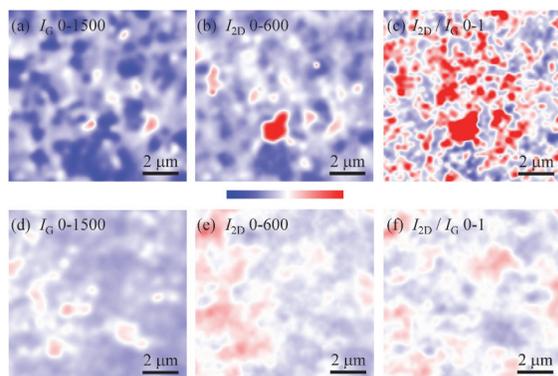

Fig. 5: (Colour on-line) (a), (b), (c) Raman intensities and intensity ratio of *G*, *2D* peak with one time transfer of 11.2 layers graphene films. (d), (e), (f) Raman intensities and intensity ratio of *G*, *2D* peak after five times stacking.

**Conclusion.** – In this letter, our numerical and experimental studies of the optical transmittance in multilayer graphene system show that the nonlinear negative exponential function $T = (1 + 1.13\pi\alpha N/2)^{-2}$ gives a good description of the light transmittance through multilayer graphene in the visible light range. It provides a simple way to determine the number of graphene layers by the measurement of the light transmittance. It is more reliable than the common method such as Raman spectroscopy, and can be generalized to other 2D materials with weak van der Waals interlayer interaction.

\*\*\*

We acknowledge the financial support from the Young Wild Idea Grant of the Delft Centre for Materials (DCMat), Delft Energy Initiative Fund and the Netherlands National Computing Facilities Foundation (NCF). This work is part of the research program of the Foundation for Fundamental Research on Matter (FOM), which is part of the Netherlands Organisation for Scientific Research (NWO).